\documentclass[reprint,
 amsmath,amssymb,
 aps,
]{revtex4-2}
\usepackage{bm, color}

\usepackage[colorlinks = true,
            allcolors = blue]{hyperref}
\usepackage{placeins}
\usepackage{float}
\usepackage{hyperref}

\usepackage{graphicx}
\usepackage{dcolumn}
\usepackage{bm}

\usepackage{printlen}

\usepackage{color,soul}
\allowdisplaybreaks

\begin{document}

\preprint{APS/123-QED}
\title{The role of effective mass and long-range interactions in the band-gap renormalization of photo-excited semiconductors}
\author{Cian C. Reeves}
\affiliation{%
Department of Physics, University of California, Santa Barbara, Santa Barbara, CA 93106
}%
\author{Scott K. Cushing}
\affiliation{%
Division of Chemistry and Chemical Engineering, California Institute of Technology, Pasadena, CA, 91125
}%
\author{Vojt\ifmmode \check{e}\else \v{e}ch Vl\ifmmode \check{c}\else \v{c}ek}
\affiliation{%
Department of Chemistry and Biochemistry, University of California, Santa Barbara, Santa Barbara, CA 93106
}%
\affiliation{%
Materials Department, University of California, Santa Barbara, Santa Barbara, CA 93106
}

\date{\today}
\begin{abstract}
Understanding how to control changes in electronic structure and related dynamical renormalizations by external driving fields is the key for understanding ultrafast spectroscopy and applications in electronics. Here we focus on the band-gap's modulation by external electric fields and uncover the effect of band dispersion on the gap renormalization. We employ the Green's function formalism using the real-time Dyson expansion to account for dynamical correlations induced by photodoping. The many-body formalism captures the dynamics of systems with long-range interactions, carrier mobility, and variable electron and hole effective mass. We also demonstrate that mean-field simulations based on the Hartree-Fock Hamiltonian, which lacks dynamical correlations,  yields a qualitatively incorrect picture of band-gap renormalization. We find the trend that increasing effective mass, thus decreasing mobility, leads to as much as a $6\%$ enhancement in band-gap renormalization. Further, the renormalization is strongly dependent on the degree of photodoping. As the screening induced by free electrons and holes effectively reduces any long-range and interband interactions for highly excited systems, we show that there is a specific turnover point with minimal band-gap. We further demonstrate that the optical gap renormalization follows the same trend though its magnitude is altered by the Moss-Burstein effect.
\end{abstract}

\maketitle

\section{\label{Intro}Introduction}

Understanding how material properties can be dynamically controlled and enhanced to create novel, tunable devices is a critical area of material science research\cite{Krausz_2009,delatorre_2021,Boschini_timeresolved_2024}. In particular, ultrafast non-equilibrium processes  are attracting great attention in recent years due to promising applications in a range of areas such as optoelectronics, metrology and quantum computation\cite{Mashiko2018,Garg2016,Schiffrin2013,Schultze2013,Krausz2014,Zong2023,clark_2007,Wang2017} and for their importance in understanding and interpreting ultrafast experiments such as time resolved angle resolved photoemission spectroscopy(TR-ARPES)\cite{Boschini_timeresolved_2024,soboto_ultrafast_2012,bovensiepen_elementary_2012,Schmitt_ultrafast_2011,bressler_ultrafast_2004,lin_carrier_2017,Liu_2023,cushing_differentiating_2019,Dong_2021,perfetti_2006}. 

Some of these applications, such as PHz switching in optoelectronics, require not only the photo-excitation of charge carriers but also the subsequent control of this non-equilibrium charge distribution. Precise control within these systems thus relies on knowledge of the properties of the material both at equilibrium and after photoexcitation.  The strong non-equilibrium typical in ultrafast experiments means that often, after excitation the properties can no longer be satisfactorily captured with an equilibrium or even linear response description. For example, upon photoexciation the band-gap and bandstructure in photoexcited gapped systems can be modified, which in turn can alter the properties of a material compared to equilibrium.  This effect, referred to as band-gap renormalization is a competition between two opposing contributions: a narrowing of the gap arising due to attraction between photoexcited electrons in the conduction band and holes in the valence band, and a widening of the gap due to the Pauli-blocking driven Burstein-Moss effect\cite{Chernikov_2015_2,Ulstrop_2016,Roth_2019,Kang_2017,Zhang_2022,Lebens_2016,yoshikawa_1999,zhang_2019,Kim_2015}. The importance of the band-gap in determining a material's properties makes the understanding of non-equilibrium band-gap renormalization crucial for leveraging the excited state properties of materials.   

In parallel to time-resolved experiments, theoretical simulations are being developed to address the non-equilibrium phenomena and predict the behavior of driven systems at time-scales associated with the redistributions of electronic populations\cite{Lian_2018,Tancogne-Dejean_2018,Schlunzen_2020,Tancogne-Dejean_2020,Beaulieu_2021,Kaye_2021,Pavlyukh_2022,granas_2022,Joost_2022,Kalvova_2024,Reeves_2024}.  Despite the theoretical advances and the importance of the phenomenon, non-equilibrium studies of band-gap renormalization are scarce, with the exception of pioneering studies which resorted to various static approximations in order to overcome high costs associated with including dynamical effects\cite{Sarma_1990,Spataru_2004,pogna_2016}; the role of the dynamical (time-non-local) correlations has, to the best of our knowledge, not been explored up to now. Here we specifically consider the non-equilibrium Green's function (NEGF) formalism, as it is directly related to experimental observations in TR-ARPES\cite{freericks2009theoreticaldescription}
and yields the dynamics of individual electronic states. In practice the widespread application of the NEGF based approach is hindered by the high cost of their time evolution. The integro-differential Kadanoff-Baym equations (KBEs), representing the NEGF equation of motion, scale cubically in the number of time steps. As a result, long-time and large-scale dynamics are intractable and overcoming the scaling bottleneck has been an active area of research. In our recent work, we have introduced the real-time Dyson expansion (RT-DE), which includes dynamical self-energy effects in the time-evolution of the NEGF while retaining effectively linear scaling in the number of time steps\cite{Reeves_2024}.  This result is particularly important for accurately and efficiently simulating time-resolved spectral properties, where one previously had to resort to the KBEs with their poor numerical scaling or static approaches that neglect dynamical many-body effects. 

We employ the RT-DE, which goes beyond previously applied static or quasi-equilibrium approximations, to simulate the TR-ARPES spectrum for a set of gapped, two-band lattice models. Using these simulated spectra we extract the band-gap renormalization as a function of the photoexcited density and uncover that band-gap renormalization is driven by the time-non-local correlations encoded in the dynamical self-energy.  These results validate the ability of the RT-DE to provide a correct description of band-gap renormalization for realistic model parameters.  Despite not having access to a theoretical benchmark our results are qualitatively consistent with experimental measurements for low photo-doped systems\cite{Chernikov_2015_2,Ulstrop_2016,Roth_2019,Kang_2017}.  Furthermore, our results show the inability of the uncorrelated time-dependent Hartree-Fock (TD-HF) approach to correctly describe band-gap renormalization, highlighting the importance of including dynamical self-energy effects. In addition to this study serving as a validation of the RT-DE we present two novel results: 1) the renormalization is strongly affected by carrier mobilities (controlled by excited carrier effective masses), with higher effective masses leading to stronger renormalization of the gap. 2) The gap can exhibit a strong non-monotonic behavior with a renormalization turnover determined by the interplay of mobility, excited carrier density and the environment dielectric screening. To our knowledge, the turnover point at higher excitation densities has not yet been observed experimentally and is only reported theoretically for quasi-equilibrium in Ref.  \cite{Spataru_2004}.  As far as we are aware ours is the first work that demonstrates that this turnover point in driven systems and that it depends on the screening of long range interactions and the carrier effective mass.  Dynamical correlations and screening are crucial in explaining both of these results as is discussed later in the manuscript, highlighting the importance of these effects included within the RT-DE.

The remainder of the paper will be organized as follows:  In section \ref{sec:Theory} we will introduce the relevant theory for this work, specifically we will discuss the model systems studied and our observables as well as briefly introducing the RT-DE.  In section \ref{sec:Methods} we discuss the different model parameters used in generating the results that are analyzed in detail in section \ref{sec:results}. Finally, in section \ref{sec:discussion} we provide an in depth discussion of the implications of our results.

\section{Theory}\label{sec:Theory}
\subsection{Time-resolved Spectral Properties}
The key observable for this study is the time and angle-resolved photoemission spectral (TR-ARPES) function, containing information on the quasiparticle (QP) states and the distribution of electrons and holes across the QP states as a function of time. Here, we use the TR-ARPES function to extract the band-gap for a range of excited, non-equilibrium populations.  The theoretical description of the TR-ARPES function is given by\cite{freericks2009theoreticaldescription}
\begin{equation}\label{eq:spectral_function}
        \begin{split}
            \mathcal{A}(\vec{k},\omega,t_p) = \int dt dt'\medspace \mathrm{e}^{-i\omega(t-t')}\mathcal{S}(t-t_p)\mathcal{S}(t'-t_p)G^<_{\vec{k}}(t,t').
        \end{split}
\end{equation}
This expression is most commonly used in the NEGF simulation of time-resolved spectra\cite{sentef_2013,Kemper_2015,Tuovinen_2020,SCHULER2021}. For simplicity the above expression approximates the matrix elements of the photoemission process (that depend on physical details of the material sample and it's surface) as constant. However, the expression is still proportional to the measured photoemission signal. $\mathcal{S}(t)$ represents the experimental probe window and determines the energy and temporal resolution of $\mathcal{A}(\vec k, \omega,t_p)$ \cite{randi2017bypassingtheenergytime}.  In the remaining we choose $\mathcal{S}(t)$ to be a Gaussian with standard deviation $\sigma$ governing the probe width. Computing equation \eqref{eq:spectral_function} requires the time-evolution of the lesser Green's function, $G^<(t,t')$, for all $t,t'$ within the probe window.

\subsection{The Real-Time Dyson Expansion} \label{sec:RTDE}
In practice, the calculation of $G(t, t')$ while including dynamical correlation effects ---which are critical for driven systems as we illustrate later--- is computationally expensive and needs to be circumvented for long time-evolutions.  Solving the formally exact Kadanoff-Baym equations (KBEs) is prohibitively expensive in the systems we study here, and we instead employ the real-time Dyson expansion (RT-DE).  The RT-DE is a recently introduced\cite{Reeves_2024} approximate solution to the integro-differential KBEs and yields the non-equal time GF.  It is equivalent to evaluating the collision integral, $I$ in equation \eqref{eq:GF_eom}, with mean-field propagators. The GF equation of motion is then recast to a set of coupled ordinary differential equations (ODEs) for $G(t,t')$ and a two-body propagator, $\mathcal{F}(t,t')$. Formally, the evolution is,
\begin{equation}\label{eq:GF_eom}
\frac{\mathrm{d}G^{\mathrm{<}}(t,t')}{\mathrm{d}t} = -i\left[h^{\mathrm{MF}}(t) G^{\mathrm{<}}(t,t') + I^{\mathrm{<}}(t,t')\right],
\end{equation}
where the collision integral is 
\begin{equation}\label{eq:coll_int}
I^{\mathrm{<}}_{im}(t,t') = -\sum_{klp}w_{iklp} \mathcal{F}_{lpmk}(t,t').
\end{equation}
The RT-DE requires a simultaneous propagation of $\mathcal{F}_{lpmk}(t,t')$, according to:
\begin{widetext}
\begin{equation}\label{eq:RT-DE}
       \begin{split}
\frac{\mathrm{d}\mathcal{F}_{lpmk}(t,t')}{\mathrm{d}t} &{=  \sum_{qrsj} (w_{qrsj} - w_{qrjs})\Big{[} G^{>,\textrm{MF}}_{lq}(t)G^{<,\textrm{MF}}_{pr}(t)G^{>,\textrm{MF}}_{sk}(t) - G^{<,\textrm{MF}}_{lq}(t)G^{<,\textrm{MF}}_{pr}(t)G^{<,\textrm{MF}}_{sk}(t)\Big{]} G^{\mathrm{<}}_{jm}(t,t')} \\&\hspace{50mm}{-i\sum_{x}\Big{[} h^\mathrm{MF}_{lx}(t) \mathcal{F}_{xpmk}(t,t') + h^\mathrm{MF}_{px}(t) \mathcal{F}_{lxmk} (t,t') - \mathcal{F}_{lpmx}(t,t') h^\mathrm{MF}_{xk}(t)\Big{]}.}\\
   \end{split}
\end{equation}
\end{widetext}
In the above expressions $w_{ijkl}$ is the bare coulomb interaction tensor, $G^\mathrm{MF}(t)$ are the mean-field propagators on top of which self-energy corrections are added,  $G^<(t,t')$ is the non-equal time electron GF within the RT-DE approximation and $\mathcal{F}(t,t')$ is a two-body propagator that contains information about many-body correlations in the system.  Equation \eqref{eq:RT-DE} is derived in Appendix \ref{sec:appendixa} and uses the second Born self-energy, which is commonly used in non-equilibrium studies due to its low-cost of implementation and relatively high accuracy. The initial conditions used here are based off of a mean-field reference GF so that, $G^<(t,t) = G^{<,\mathrm{MF}}(t)$ and $\mathcal{F}(t,t) = 0$.  Thus the initial populations and population dynamics are given at the mean-field level and do not include dynamical self-energy effects.   This analogous to the commonly applied equilibrium one-shot self-energy correction\cite{godby_1986,godby_1988,Chelikowsky1996QuantumTO,AULBUR2000,rohlfing_2000,golze_2019,Klein_2023}, where an equilibrium mean-field GF is used as a reference compared to the non-equilibrium mean-field GF used in the RT-DE.  In principle other initial conditions are possible, such as one taken from the HF-GKBA, however this has yet to be fully explored.

The RT-DE scales effectively linearly in the number of time-steps, making it an efficient method for including dynamical correlations in non-equilibrium spectral properties in particular when compared to the Kadanoff-Baym equations (which scale cubically with the number of timesteps). In essence, the RT-DE is equivalent to the one-shot many-body perturbation theory correction on top of a mean-field density matrix, and it naturally reduces to the conventional one shot perturbative corrections when expanded around the ground state reduced one-body density matrix \cite{Reeves_2024}.  In this work we use it to qualitatively study the effect of dielectric screening, carrier mobility and many-body correlations in phot-induced band-gap renormalization of semiconductors and insulators.  For a more detailed discussion of the RT-DE, including benchmarks and derivations of the equations of motion, we direct the reader to Ref. \cite{Reeves_2024}.

\subsection{Model System}
To study the role effective carrier mass plays in band-gap renormalization we choose a periodic Hamiltonian with two-bands and long range density density interactions.  We use a high $k$-point resolution (16 $k$-points in this work) on an effective one-dimensional model :
\begin{widetext}
 \begin{equation}\label{eq:multiband_ham}
\begin{split}
        \mathcal{H} &= \sum_{\alpha,\beta\in \{c,v\}\atop{i,j,\sigma}}h^{\alpha\beta}_{ij}(t)c^{\dagger\alpha}_{i,\sigma}c^{\beta}_{j,\sigma} + U_\mathrm{intra}\Bigg{[}\sum_{i,\alpha}n^\alpha_{i\uparrow}n^{\alpha}_{i\downarrow} +\sum_{i < j,\alpha} \frac{n_{i}^\alpha n_{j}^\alpha}{\varepsilon|\Vec{r}_i - \Vec{r}_j|}\Bigg{]}+U_\mathrm{inter}\Bigg{[}\sum_{i}n_{i}^cn_{i}^v + \sum_{i < j} \frac{n_{i}^c n_{j}^v}{\varepsilon|\Vec{r}_i - \Vec{r}_j|}\Bigg{]}.\\
        h^{\alpha\beta}_{ij}(t) &= \underbrace{J_\alpha\delta_{\alpha\beta}\delta_{\langle i,j\rangle}+ \epsilon_\alpha\delta_{\alpha\beta} }_{h^{(0)}}+ \underbrace{\delta_{ij}(1-\delta_{\alpha\beta})
        E\cos(\omega_p(t-t_0)) \mathrm{e}^{-\frac{(t-t_0)^2}{2T_p^2}}}_{h^{\mathrm{drive}}(t)}.
\end{split}
\end{equation}   
\end{widetext}
Here $h^{(0)}$ describes the kinetic energy and onsite portion of the single particle Hamiltonian and $h^{\mathrm{drive}}(t)$ describes an optical excitation between the conduction and valence band that drives the system from equilibrium. $U_\mathrm{intra}$ and $U_\mathrm{inter}$ determine the strength of the coulomb interaction within and between the orbitals of the model.   Throughout this study we are interested in how the measured band-gap renormalization is altered by changing the effective mass of carriers in the system (governed by $ J_\alpha^{-1}$) and the extrinsic static dielectric screening ($\varepsilon$) which determines the attenuation of the intra and interband interactions. Note that the dynamical renormalization of the two-particle terms is fully taken into account, so the value $\epsilon $ effectively interpolates between local and non-local interaction models, and it is interpreted as the effect of an external dielectric environment.

\section{Methods}\label{sec:Methods}
For the Hamiltonian in equation \eqref{eq:multiband_ham} we use $N_k = 16$, which ensures all our calculations are converged with respect to the system size. The system is at half-filling (i.e., the valence band is fully filled, while the conduction band is completely empty).  The calculations are carried out at zero temperature. We fix the valence band hopping parameter, $J_v = 1$ and use this as our energy unit.  To study how effective mass affects the band-gap renormalization, we perform simulations for the following values of the conduction band hopping (equivalent to different the effective masses, see Fig. \ref{fig:bandstructure}) $J_c \in \{0.2,0.4,0.6,1.0,1.4,1.8\}$. These values of the hopping strengths and corresponding effective masses have been chosen to encompass a range that is observed in typical semiconductor materials for the ratio of conduction and valence band effective masses\cite{green_1990,harrison1989electronic,sze2006physics}.
 For the results shown in sections \ref{sec:renormalization} and \ref{sec:BM_shfit} we fix $U_\mathrm{intra} = 3.0J_v$ and $U_\mathrm{inter}=0.5J_v$. We have chosen $U_\mathrm{inter}\ll U_\mathrm{intra}$ to reduce excitonic effects in the system.  The onsite potentials $\epsilon_\alpha$ are chosen for each set of parameters such that the chemical potential sits at zero and so that the equilibrium gap is fixed to $E_g = 5J_v$.  As we will see in section \ref{sec:BM_shfit} this value of $E_g$ and the values of $U_{\mathrm{inter/intra}}$ leads to an exciton binding energy about $1\%$ of the total gap.  This is within the range occurring for typical semiconductors justifying this a physically reasonable choice of parameters\cite{Boer2018}.
Finally we vary $\varepsilon \in \{5,10,\infty\}$ to see how dielectric screening changes the band-gap renormalization properties of the material. These values are chosen to encompass a range of material environments, $\varepsilon = \infty$ corresponding to the extreme case of a metallic environment while $\varepsilon=10$ and $\varepsilon=5$ corresponds to values in semiconducting/insulating systems such as silicon or hexagonal boron nitride respectively         \cite{2007introduction,Dunalp_1953,Laturia2018}

The time dependent pulse in equation \eqref{eq:multiband_ham} is used to excite different populations of electrons to the conduction band.  For the results shown in this manuscript we fix $T_p = 2.0J_v^{-1}$ and $t_0 = 10J_v^{-1}$. We pump the system at a frequency above the gap, in the case of onsite only interactions ($\varepsilon = \infty$) we have $\omega_p = 5.2J_v$, while for the cases with $\varepsilon = 5,10$ we have $\omega_p = 5.3J_v$.  The higher $\omega_p$ for finite $\varepsilon$ is related to the fact that we use the TD-HF density matrix as a starting point for the RT-DE.  As we will see in section \ref{sec:results} the TD-HF gap opens more for $\varepsilon=10,5$ and so we must pump further above the gap in order to reach the more highly excited populations. 
The field strength parameter, $E$, is varied in order to get a range of roughly equally spaced excited populations between $n_c \approx 0.1\%$ and $ 7\%$. This ensures we cover a range of experimentally achievable values from medium/low up to what is considered relatively high excitation densities\cite{Hedayat_2021,Spataru_2004,CALLAN2000167}. 
Tables 1-3 in the supplemental material show the exact values of $E$ used for each model considered\cite{supp}.  The standard deviation of the Gaussian windowing function in equation \eqref{eq:spectral_function} is chosen to be $\sigma = 8J_v^{-1}$.  Scaling our gap of $E_g = 5J_v^{-1}$ to a typical semiconductor gap of around $.1-5\mathrm{eV}$\cite{2007introduction,streetman2000solid} we can perform a unit analysis that tells us $1J_v^{-1}\sim 1\mathrm{fs}-50\mathrm{fs}$ making this a experimentally achievable and thus physically reasonable choice of parameter\cite{Boschini_timeresolved_2024,soboto_ultrafast_2012,Huang2022,smallwood_2014,schmitt_2008,Schmitt_ultrafast_2011}. 

Finally, the RT-DE requires that we first perform a time-evolution of the equal-time GF with a mean-field Hamiltonian.  For the results shown here we have chosen TD-HF as our starting point, which is equivalent to time evolving equation \eqref{eq:GF_eom} with $I(t,t') =0$ and $h^\mathrm{MF}(t) = h^\mathrm{HF}(t)$. We have used fourth order Runge-Kutte to propagate the equations of motion with a time-step of $dt = 0.05J_c^{-1}$ for both the TD-HF and RT-DE time evolutions.

\section{Results}\label{sec:results}
\begin{figure}
    \centering
    \includegraphics[width=.5\textwidth]{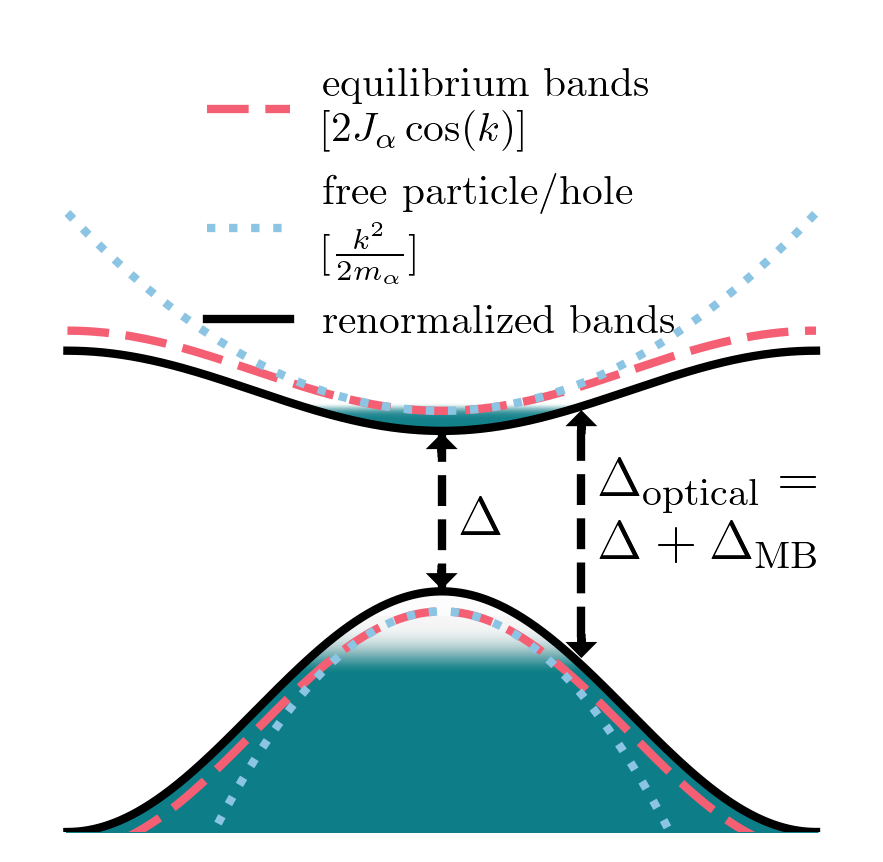}
    \caption{Schematic of the bandstructure and extracted values studied in sections \ref{sec:renormalization} and \ref{sec:BM_shfit}.  Nearest neighbor hopping results in a cosine band shape. The conduction band effective mass is tuned through the hopping parameter $J_c$ ($J_c\approx \frac{1}{2m_c}$ near the gamma point).  Upon excitation the equilibrium bands are shifted (pink dashed lines moving to solid black lines).  Section \ref{sec:renormalization} deals with the renormalization of the distance from conduction band minima to valence band maxima, denoted by $\Delta$.  Section \ref{sec:BM_shfit} deals with the renormalization of the optical gap, denoted by $\Delta_\mathrm{optical}$, that includes the shift of the bands as well as the Burstein-Moss shift, brought about by Pauli-blocking in the conduction band.}
    \label{fig:bandstructure}
\end{figure}

We will now present the results for non-equilibrium simulations that reveal the relation between the effective mass, dielectric screening, and the magnitude of the  band-gap renormalization.  The results we present in sections \ref{sec:renormalization} and \ref{sec:BM_shfit} are taken from band structures produced with the RT-DE and TD-HF GF dynamics.  The reported quantities are shown schematically in Fig. \ref{fig:bandstructure}.  Section \ref{sec:renormalization} deals with the renormalization of the quantity $\Delta$ which measures the distance between the conduction and valence band extrema, while sectioin \ref{sec:BM_shfit} deals with the shift in the optical gap, $\Delta_\mathrm{optical}$, which includes contributions from $\Delta$ as well as a positive contribution from the Burstein-Moss shift.  

\subsection{Role of effective mass and dielectric screening in band-gap renormalization}\label{sec:renormalization}
\begin{figure*}
    \centering
\includegraphics[width=7.0833in]{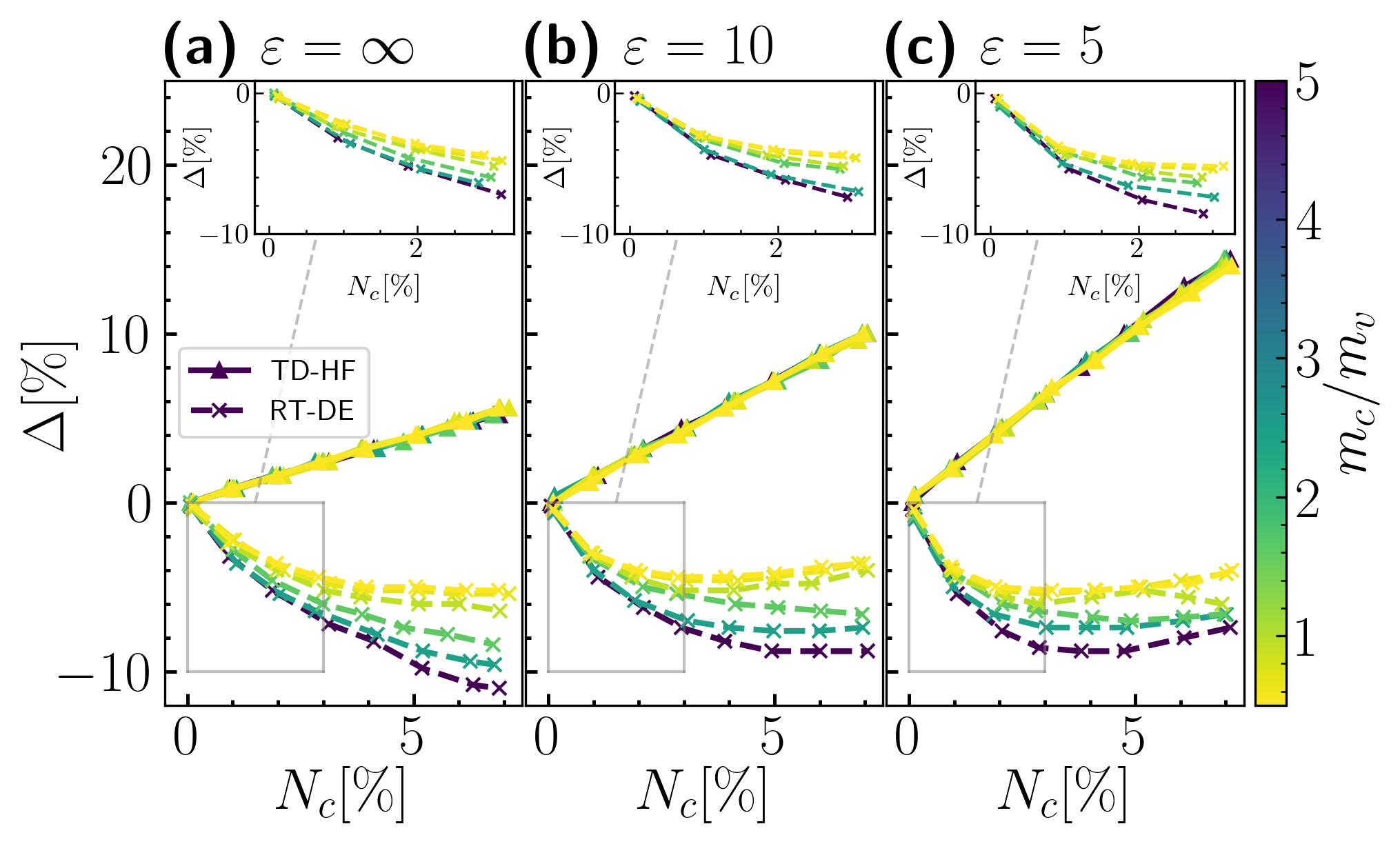}
    \caption{Percent renormalization of the gap ---as measured from the conduction band minimum to the valence band maximum --- as a function of excited carrier density, $N_c$, for the Hamiltonian given in equation \eqref{eq:multiband_ham} computed with TD-HF (solid triangles) and RT-DE (dashed crosses).  The renormalization is also shown as a function of the conduction/valence band effective mass ratio, $m_c/m_v$ and for different values of the dielectric screening constant: a) $\varepsilon=\infty$, b)$\varepsilon=10$, and c) $\varepsilon=5$. Insets show the renormalization computed with RT-DE zoomed in around $-10\%<\Delta<0\%$ and $0\%<N_c<3\%$}
    \label{fig:BGR}
\end{figure*}

In this section, we inspect the gap changes, i.e., the variation of the distance between the maximum of the valence band and the minimum of the conduction band. The energy gap is illustrated in Fig. \ref{fig:bandstructure} where we distinguish it from the optical gap which is modified by the Pauli blocking upon photoexcitation (i.e., the Burstein-Moss shift discussed in the next section). The figure also shows the free particle/hole dispersion around the band edge determined by the effective mass and controlled by the hopping parameter in our model (see section \ref{sec:Methods}). 

The results are summarized in Fig. \ref{fig:BGR}  showing the percentage renormalization of the band gap, $\Delta$, as a function of the percent of photoexcited carriers, $N_c$. Different panels represent results for distinct values of the interaction range, characterized by the environment dielectric screening constant, $\varepsilon$. Further, we employed models with distinct conduction-valence effective mass ratios, $m_c/m_v$ showed in the figure by distinct color.  The specific parameters used are given in section \ref{sec:Methods}.  We compare results given by TD-HF and the RT-DE using the second Born self-energy. 

We first inspect the TD-HF results for $\varepsilon=\infty$, shown in Fig. \ref{fig:BGR} a),  where TD-HF refers to the time-evolution using equation \eqref{eq:GF_eom} with $I(t,t') =0$ and $h^\mathrm{MF}(t) = h^\mathrm{HF}(t)$. We see a linear increase of the gap with the carrier density; the observed slope is independent of the effective mass. The renormalization reaches $\Delta = +5.2\%$ at $N_c\approx7\%$.  The TD-HF result exhibits the same linear behavior when long-range interactions are introduced by making the dielectric constant, $\varepsilon$, finite.   The only difference is an increase in the slope with the range, i.e., for $\varepsilon<\infty$.  This results in the renormalization at $N_c\approx 7\%$ increasing to $+10\%$ for $\varepsilon=10$ and $+14\%$ for $\varepsilon=5$.

Moving now to the RT-DE results, which include dynamical correlations at the level of the second Born approximation, we see two notable differences. \textit{First}, the gap change is indeed dependent on the effective mass (or the ratio between the particle and hole effective masses).  For higher $m_c/m_v$ ratios, the band-gap renormalization increases.  This is a persistent trend that appears for all three dielectric screening values and is present even at low excitation densities as shown in the inserts of Fig. \ref{fig:BGR} (a)-(c). The insets illustrate the weak photodoping limit, readily accessible by experiments. At $N_c\approx3\%$ changing the effective mass ratio from $m_c/m_v=5$ to $m_c/m_v=0.556$ leads to  $\Delta_{\mathrm{heavy-light}}\approx 2.8\%,3\%$ and $3.4\%$ for $\varepsilon=\infty,10$ and $5$ respectively.  While at $N_c\approx7\%$ we have $\Delta_{\mathrm{heavy-light}}\approx 5.8\%, 5.2\%$ and $3.4\%$ again for $\varepsilon=\infty,10$ and $5$ respectively. 

The \textit{second} important difference is a strongly non-linear behavior in response to the photodoping strength (i.e., as a function of $N_c$). For all the cases studied the RT-DE initially narrows the gap followed by --- depending on the effective mass and environment dielectric screening --- a reopening of the gap, which is in strong contrast to the linear monotonic behavior of TD-HF.  For $\varepsilon=\infty$ in Fig. \ref{fig:BGR} (a), and the heaviest mass ratio $(m_c/m_v = 5)$ the gap is renormalized by around $-11\%$ at $N_c\approx7\%$ and shows no inflection point at which the gap begins opening again.  For the same value of $\varepsilon$ the lightest mass ratio $(m_c/m_v=0.556)$ the renormalization only reaches around $-5\%$ for $N_c\approx7\%$.  Furthermore, for the same mass ratio  $\Delta$ remains roughly constant after $N_c$ reaches $\approx5\%$. Including the long range interactions leads to an interesting change in the renormalization behavior.  Despite the long range interacting models initially having a stronger renormalization of the gap --- reaching a maximum at $N_c\approx3\%$ of $\Delta=-8.6\%$ for $\varepsilon=5$ compared to $\Delta=-7.2\%$ for $\varepsilon=\infty$  --- they ultimately change the gap less for higher excitation densities.  For $\varepsilon=10$ we see the mass ratio $m_c/m_v=5$ begins to plateau around $N_c=5\%$, where it had shown no clear sign of flattening for $\varepsilon=\infty$.  Additionally, for the lightest mass ratio $(m_c/m_v=0.556)$ we see that the gap, after initially narrowing, begins to re-open at around $N_c=4\%$.  This trend continues as we further decrease $\varepsilon$ in Fig. \ref{fig:BGR} (c).  Even for the heaviest mass ratio, $m_c/m_v=5$, we see an inflection point at $N_c\approx4\%$ with a similar upturn occuring for the other effective mass ratios studied.

\subsection{Burstein-Moss Shift}\label{sec:BM_shfit}
This section we analyzes the renormalization of the optical band-gap which is accessible by optical absorption spectroscopies\cite{makula_2018,DOLGONOS201643}. For the systems studied here (with the parameters selected to match the behaviors of relevant semiconductors) the excitonic signatures are absent (due to the weak electron-hole couplings),  with the exception of the systems at the extremely low photodoping discussed below. The details of how we compute the optical gap renormalization are given in the supplemental material\cite{supp}.  Fig. \ref{fig:Burstein_moss} shows the renormalization of the optical band gap as well as the Burstein-Moss shift as a function of the percentage of photoexcited carriers, $N_c$.  The results shown in Fig. \ref{fig:Burstein_moss} are for the onsite interacting model ($\varepsilon=\infty$) with conduction band hopping strengths, $J_c=0.2J_v$, $0.6J_v$, $1.0J_v$, $1.4J_v$ and $1.8J_v$.  In this section we analyze the RT-DE results only.

Clearly, the optical gap, $\Delta_{\rm opt}$ shows a similar dependence on the effective mass ratio compared to that seen in the previous section, exhibiting the same trend of increased renormalization with increasing effective mass. The primary difference between the renormalization of the optical gap compared to the gap studied in section \ref{sec:renormalization} comes from  the Burstein-Moss shift, which increases the apparent gap due to Pauli blocking\cite{Burstein_1954,Moss_1954}.  Thus, the overall change in $\Delta_{\rm opt}$ caused by photodoping will combine the renormalization $\Delta$ as well as the Burstein-Moss effect.   For $m_c/m_v=1.0$, $0.714$ and $0.556$, this positive shift causes $\Delta_{\rm opt} >0$, despite $\Delta<0$ as measured between the band  extrema, see Fig. \ref{fig:BGR} (a).  For the heavier mass ratios of $m_c/m_v=5$ and $1.667$ the overall shift of the bands towards the Fermi energy dominates over the Burstein-Moss effect and leads to the optical band gap shrinking. The dependence on effective mass is apparently stronger for the optical gap than for the gap in section \ref{sec:renormalization}, where $\Delta_{\mathrm{heavy-light}}\approx 9\%$ compared to the  $6\%$ observed in the onsite model in the previous section.

A final noteworthy observation in these results is the behavior of the optical gap as $N_c\to 0\%$.  As the photodoping approaches the linear response regime, the renormalization for different masses appear to converge to a common value.  They do not however converge to 0 but rather some finite $\Delta$, as measured relative to the initial distance between valence and conduction band extrema $E_g=5J_v$, of around $\Delta=1\%$.  This is consistent with weak a excitonic coupling giving rise to a finite difference between the optical and fundamental gap in the linear response regime.  This is discussed further in section \ref{sec:discussion}.

\begin{figure*}
    \centering
    \includegraphics[width=\textwidth]{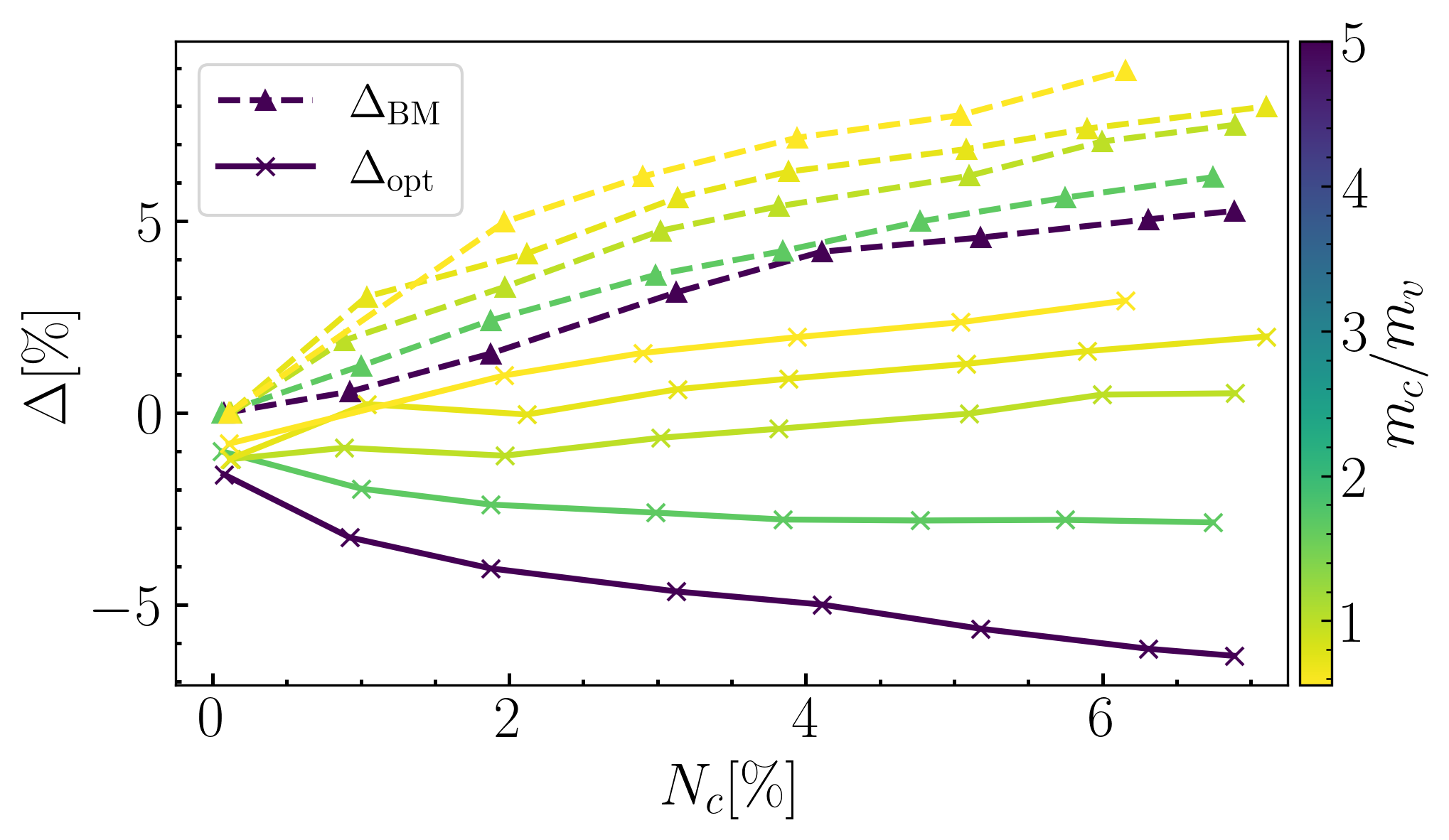}
    \caption{Optical band gap(solid crosses) shift and Burstein-Moss shift(dashed triangles) measured relative to the fundamental gap, $E_g=5J_v$, as a function of excited carrier density and computed for the onsite interacting model.}
    \label{fig:Burstein_moss}
\end{figure*}
\section{Discussion and Conclusions}\label{sec:discussion}

The strong effect of the dynamical correlations warrants further discussion, also in the light of possible band-gap renormalization tunability through effective mass manipulation. We have investigated strongly photodoped systems with TD-HF and the RT-DE and observed two striking differences between the results produced by each method. 

First, we will discuss the sign of the renormalization.  For the parameters chosen, TD-HF shows a nearly linear increase in the band-gap as a function of the photo-excited density.  This is contrary to what is observed experimentally\cite{Chernikov_2015_2,Ulstrop_2016,Roth_2019,Kang_2017}, where the band gap typically narrows due to the attraction between electrons in the conduction band and holes in the valence band.  We note that while this is true for the minimum gap between the conduction and valence states, as we have seen in Sec. \ref{sec:BM_shfit} this is not necessarily the case for the optical band gap.  TD-HF cannot capture even qualitative features of the band-gap changes. In detail, the Coulombic (Hartree) term dominates the renormalization in TD-HF and as we show in the supplemental material \cite{supp} for the model parameters in this paper this term causes the band-gap to widen.  Conversely, the RT-DE shows non-linear behavior and, more importantly, causes an initial decrease in the band gap for low excitation densities.  The introduction of a dynamical self-energy also implies the introduction of screened quasiparticles and their interactions explicitly incorporated to the two-particle correlator $\mathcal{F}$ entering the collision integral in Eq.~\ref{eq:coll_int}.  These excited quasi-electrons and quasi-holes attract one another, and for low to medium excitation densities, this attraction dominates over the Hartree term, thus leading to a shrinkage of the band gap.  We also observe that for some model parameters and for higher excitation densities the $\Delta$ starts to behave  differently and the band-gap begins to open again.  This has been observed in quasi-equilibrium results in Ref. \cite{Spataru_2004} for GaAs, however to our knowledge has not been seen in experiment.  That this upturn occurs for high excitation densities, we attribute to screening from many-body effects.  As more electrons are excited to the conduction band they will screen the quasiparticle attraction, allowing the Hartree term to take over the dominant role in renormalizing the band leading to $\Delta$ increase. This is also supported by the fact that as the magnitude of long range interactions is increased  (for low $\varepsilon$) the inflection point is reached at smaller $N_c$, since the long range interactions tend couple to density fluctuations and hence screen more intensely.  In contrast to a local model, i.e., $\varepsilon\to \infty$, each electron can have an effect on a greater portion of the system thus screening more effectively. Currently, these changes in $\Delta$ seem appear only at high photodoping, which may be observed experimentally only on very short time scales (before radiation damage occurs). However, the extrinsic dielectric environment (governing the interaction range) possibly allows for this turnover to be observed at experimentally achievable excitation densities and may thus serve as a control knob.

The second obvious difference between the RT-DE and TD-HF results is how changing effective mass affects the band-gap renormalization.  TD-HF shows practically no relationship $\Delta$ and the effective mass while the RT-DE shows a clear positive correlation between those quantities. This result can be explained by the relationship between effective mass and band localization in real-space.  In principle, quasiparticles with larger effective masses are effectively more localized, i.e., less mobile. Because TD-HF is a mean-field method and averages over all carriers it does not reflect the notion of band localization or quasiparticle mobility.  We thus observe no relationship between the effective mass and band-gap renormalization for a mean-field TD-HF approaches.  

The positive correlation between effective mass and renormalization for the RT-DE represents one of the major results of this paper.  Here, the introduction of screening entering the dynamical self-energy is crucial.    Large effective mass is associated with lower mobility and generally less pronounced screening (due to charge density fluctuations) of the electron-hole attraction that governs $\Delta$. The reduced bandwidth then translates to a stronger change of $\Delta$.   Conversely, light carriers are responsible for strong screening, which is especially true for strong photodoping, leading to a large number of light (and thus polarizable) electrons. This then reduces the attraction between the free quasi-electrons and quasi-holes and ``screens out'' the gap decrease.  This argument also explains why, in the RT-DE, the maximal gap renormalization depends on the effective mass: $\Delta$ stops decreasing first for the lowest effective masses. This gap reopening is again due to the screening of the electron-hole attraction, enabling the Hartree term to dominate the renormalization; this descreening happens more effectively for electrons with a low effective mass.

Finally, Sec. \ref{sec:BM_shfit} presents the results for the renormalization of the optical band gap and the Burstein-Moss shift.  This shows that the role effective mass plays in the renormalization of the gap is not confined to the valence (conduction) band maximum (minimum), and that it should be observable in both photoemission and optical absorption spectroscopy experiments.  The small non-zero value of the optical gap for $N_c\rightarrow 0$ can be explained by the small interband interaction included in our model. This gives rise to spectral features that are consistent with a weak exciton coupling.  Indeed, the results in Fig. \ref{fig:Burstein_moss} is a percent difference from the fixed fundamental gap $E_g = 5J_v$, and the $N_c\rightarrow 0$  limit represents the linear response optical gap for our model.  Further evidence for this is in the supplemental material where we computing $\Delta_\mathrm{opt}$ at small $N_c$ for increasing $U_\mathrm{inter}$ and show corresponding decrease in the linear response optical gap.  A more in depth investigation into excitonic properties within the RT-DE framework is left to a later paper.

Two implications of the results presented here potentially serve as control parameters for tuning band gap renormalization properties in materials.  Firstly, in \cite{Spataru_2004}, it was predicted for GaAs that before the band-gap could be fully closed by photo-excitation, the Hartree term would cause it to begin opening again. Here we observe similar behavior, however in addition we see that the point at which this occurs depends, at least in part, on the range of interactions.  Thus, systems with highly localized interactions may be good candidates for materials in which a full closure of the band-gap can be induced, and further, the use of different dielectric substrates can be used as an external control of the band-gap behavior under photoexcitation. The second useful control is through doping and alloying, which directly modulates the effective mass of carriers in a material and hence the strength of band-gap renormalization.  

The RT-DE results analyzed and discussed here show the importance of many-body effects for qualitatively correct predictions of band-gap renormalization.  Further, they have allowed us to observe and explain possible band-gap manipulation strategies in materials.  Several extensions to this study are technically challenging, and they will be pursued in future works. Most importantly, it is necessary to apply the RT-DE to realistic materials, which represents a major methodological development that is underway.

\section*{Acknowledgements}
This material is based upon work supported by the U.S. Department of Energy, Office of Science, Office of Advanced Scientific Computing Research and Office of Basic Energy Sciences, Scientific Discovery through Advanced Computing (SciDAC) program under Award Number DE-SC0022198. This research used resources of the National Energy Research Scientific Computing Center, a DOE Office of Science User Facility supported by the Office of Science of the U.S. Department of Energy under Contract No. DE-AC02-05CH11231 using NERSC Award No. BES-ERCAP0032056.  Reeves is supported by the National Science Foundation through Enabling Quantum Leap: Convergent Accelerated Discovery Foundries for Quantum Materials Science,
Engineering and Information (Q-AMASE-i) award number
DMR-1906325. Cushing is supported by the Liquid Sunlight Alliance, which is supported by the U.S. Department of Energy, Office of Science, Office of Basic Energy Sciences, Fuels from Sunlight Hub under Award Number DE-SC0021266. 
\appendix
\section{Derivation of the RT-DE equations of motion}\label{sec:appendixa}
We will now outline the steps involved in deriving the equation of motion in equation \eqref{eq:RT-DE}.   Here we apply the second-Born self-energy but our scheme is generalizable to other components of the GF as well as other self-energy approximations including $GW$\cite{Reeves_2024}. The lesser component of the collision integral has the following explicit expression,
\begin{equation*}
    \begin{split}
        I^<(t,t') &= \int_{0}^t d\Bar{t}\medspace\Sigma^{\mathrm{R}}(t,\Bar{t})G^{<}(\Bar{t},t') + \int_0^{t'}d\Bar{t}\medspace\Sigma^{>}(t,\Bar{t})G^{\mathrm{A}}(\Bar{t},t').\\
    \end{split}
\end{equation*}
For the second-Born self-energy $\Sigma^{\mathrm{R}}(t,t')$ and $\Sigma^<(t,t')$ are given by
\begin{equation}
    \begin{split}
        \Sigma^{\mathrm{R}}_{ij}(t,t') &= -\sum_{klpqrs} w_{iklp}w_{qrsj}^\mathrm{x}\bigg{[}G^>_{lq}(t,t')G^>_{pr}(t,t')G^<_{sk}(t',t) 
        \\
        &- G^<_{lq}(t,t')G^<_{pr}(t,t')G^>_{sk}(t',t)\bigg{]},\\
        \Sigma^{>}_{ij}(t,t') &= -\sum_{klpqrs} w_{iklp}w_{qrsj}^\mathrm{x}G^>_{lq}(t,t')G^>_{pr}(t,t')G^<_{sk}(t',t) 
        ,\\
    \end{split}
\end{equation}
where $w_{qrsj}^\mathrm{x} = w_{qrsj} - w_{qrjs}$ encodes the direct and exchange portion of the second-Born self-energy. Approximating the self-energy with mean-field GFs, ie $\Sigma[G]\approx \Sigma[G^\mathrm{MF}]$, leads to the following
\begin{equation*}
    \begin{split}
     &\Sigma^{\mathrm{R}}_{ij}[G^{\mathrm{MF}}(t,t')] = \\&-\sum_{klpqrs} w_{iklp}w_{qrsj}^\mathrm{x}\bigg{[}G^{>,\mathrm{MF}}_{lq}(t,t')G^{>,\mathrm{MF}}_{pr}(t,t')G^{<,\mathrm{MF}}_{sk}(t',t) 
        \\&\hspace{20mm}- G^{<,\mathrm{MF}}_{lq}(t,t')G^{<,\mathrm{MF}}_{pr}(t,t')G^{>,\mathrm{MF}}_{sk}(t',t)\bigg{]},\\
        &\Sigma^{>}_{ij}[G^{\mathrm{MF}}(t,t')] = \\&-\sum_{klpqrs} w_{iklp}w_{qrsj}^\mathrm{x}G^{>,\mathrm{MF}}_{lq}(t,t')G^{>,\mathrm{MF}}_{pr}(t,t')G^{<,\mathrm{MF}}_{sk}(t',t).\\
    \end{split}
\end{equation*}
Combining this with the expression for $I^<(t,t')$ and the relation in equation \eqref{eq:coll_int}
we have the following form of $\mathcal{F}(t,t')$ for the second-Born self-energy,
\begin{equation}\label{eq:explicit_F_form}
    \begin{split}
    &\mathcal{F}_{lpmk}(t,t') = \\&\int_{0}^t d\Bar{t}\medspace w_{qrsj}^\mathrm{x}\bigg{[}G^{>,\mathrm{MF}}_{lq}(t,\Bar{t})G^{>,\mathrm{MF}}_{pr}(t,\Bar{t})G^{<,\mathrm{MF}}_{sk}(\Bar{t},t) 
        \\&- G^{<,\mathrm{MF}}_{lq}(t,\Bar{t})G^{<,\mathrm{MF}}_{pr}(t,\Bar{t})G^{>,\mathrm{MF}}_{sk}(\Bar{t},t)\bigg{]}G^{<}_{jm}(\Bar{t},t')\\
        &+\int_{0}^{t'} d\Bar{t}\medspace w_{qrsj}^\mathrm{x}G^{>,\mathrm{MF}}_{lq}(t,\Bar{t})G^{>,\mathrm{MF}}_{pr}(t,\Bar{t})G^{<,\mathrm{MF}}_{sk}(\Bar{t},t)G^{\mathrm{A}}_{jm}(\Bar{t},t').
    \end{split}
\end{equation}
To derive the equations of motion for $\mathcal{F}(t,t')$ we take the derivative of both sides with respect to $t$.  Taking the time derivative of the right hand side with respect to the integral's time argument (denoted as $\int$ in the subscript) gives,

\begin{equation*}
 \begin{split}
    &\left[\frac{\mathrm{d}\mathcal{F}_{lpmk}(t,t')}{\mathrm{d}t}\right]_{\int}\\ &= \sum_{qrsj} w_{qrsj}^\mathrm{x}(t)\Big[ G^{>,\mathrm{MF}}_{lq}(t)G^{>,\mathrm{MF}}_{pr}(t)G^{<,\mathrm{MF}}_{sk}(t) \\&- G^{<,\mathrm{MF}}_{lq}(t)G^{<,\mathrm{MF}}_{pr}(t)G^{>,\mathrm{MF}}_{sk}(t)\Big{]} G^{\mathrm{R}}_{jm}(t,t').
 \end{split}
\end{equation*}

Taking the derivative of the $t$ dependent mean-field GFs appearing in equation \eqref{eq:explicit_F_form} (denoted as $G^\mathrm{MF}$ in the subscript) and making use of the special form of the mean-field GF equations of motion, 
\begin{equation}\label{eq:MF_eom}
    \begin{split}
        \frac{\mathrm{d}G^{\lessgtr,\textrm{MF}}_{ij}(t,t')}{\mathrm{d}t} &= -i\sum_k h^{\mathrm{MF}}_{ik}(t)G^{\lessgtr,\textrm{MF}}_{kj}(t,t'),\\
         \frac{\mathrm{d}G^{\lessgtr,\textrm{MF}}_{ij}(t',t)}{\mathrm{d}t} &= i\sum_kG^{\lessgtr,\textrm{MF}}_{ik}(t',t)h^{\mathrm{MF}}_{kj}(t),\\
    \end{split}
\end{equation}
leads to the following
\begin{equation*}
    \begin{split}
&\left[\frac{\mathrm{d}\mathcal{F}_{lpmk}(t,t')}{\mathrm{d}t}\right]_{G^\mathrm{MF}} = -i\sum_{x}\Big{[} h^\mathrm{MF}_{lx}(t) \mathcal{F}_{xpmk}(t,t')\\& + h^\mathrm{MF}_{px}(t) \mathcal{F}_{lxmk} (t,t')-   \mathcal{F}_{lpmx}(t,t') h^\mathrm{MF}_{xk}(t)\Big{]}.
    \end{split}
\end{equation*}
These are combined to give the equation of motion for $\mathcal{F}$ within the second-Born self-energy approximation as written in equation \eqref{eq:RT-DE}.

The RT-DE proceeds with an initial time evolution of $G^\mathrm{\lessgtr,MF}(t,t)$ using 
\begin{equation}\label{eq:diagonal_eom}
    \frac{dG^\mathrm{\lessgtr,MF}(t)}{dt} = -i [G^\mathrm{\lessgtr,MF}(t),h^{\mathrm{MF}}(t)],
\end{equation}
for some $h^{\textrm{MF}}(t)$. We emphasize this can be any static single particle Hamiltonian such that the time-evolution of $G^{\lessgtr,\textrm{MF}}$ follows the form of equations \eqref{eq:MF_eom} and \eqref{eq:diagonal_eom}.  Using this information for the time diagonal components in equation \eqref{eq:RT-DE} we can simultaneously propagate $\mathcal{F}(t,t')$ and $G(t,t')$ in the $t$ variable in the range $t'<t<T_{\mathrm{max}}$ with the following initial conditions for each $t'$,
\begin{equation*}
\begin{split}
        G^{<}(t',t') &= G^{<,\mathrm{MF}}(t',t'),\\
        \mathcal{F}(t',t') &= 0.
\end{split}
\end{equation*}
We refer the reader to Ref.\cite{Reeves_2024} for more detailed derivation and discussion, as well as derivations for the $GW$ self-energy approximation.  
\newpage
\bibliography{Bib}
\end{document}


\preprint{APS/123-QED}
\title{Supplementary Information for  ``The role of effective mass and long-range interactions in the band-gap renormalization of photo-excited semiconductors''}

\author{Cian C. Reeves}
\affiliation{%
Department of Physics, University of California, Santa Barbara, Santa Barbara, CA 93106
}%
\author{Scott K. Cushing}
\affiliation{%
Division of Chemistry and Chemical Engineering, California Institute of Technology, Pasadena, CA, 91125
\looseness=-1}%
\author{Vojt\ifmmode \check{e}\else \v{e}ch Vl\ifmmode \check{c}\else \v{c}ek}
\affiliation{%
Department of Chemistry and Biochemistry, University of California, Santa Barbara, Santa Barbara, CA 93106
}%
\affiliation{%
Department of Materials, University of California, Santa Barbara, Santa Barbara, CA 93106
}

\date{\today}
\begin{abstract}

\end{abstract}

\maketitle

\section{Electric Field Values}
\begin{table}[h]
    \centering
    \begin{tabular}{||c||c|c|c|c|c|c|c|c||}
    \hline
         &$E_1$&$E_2$&$E_3$&$E_4$&$E_5$&$E_6$&$E_7$&$E_8$\\
         \hline
         $J_c=0.2$& 0.02& 0.07& 0.10&0.13&0.15&0.17&0.19&0.20\\
         \hline
         $J_c=0.4$& 0.02& 0.08& 0.11&0.13&0.16&0.18&0.20&0.21\\
         \hline
         $J_c=0.6$& 0.02& 0.08& 0.11&0.14&0.16&0.18&0.20&0.22\\
         \hline
         $J_c=1.0$& 0.03& 0.08& 0.12&0.15&0.17&0.20&0.22&0.24\\
         \hline
         $J_c=1.4$& 0.03& 0.09& 0.13&0.16&0.18&0.21&0.23&0.26\\
         \hline
         $J_c=1.8$& 0.03& 0.09& 0.13&0.16&0.19&0.22&0.25&0.27\\
         \hline
    \end{tabular}
    \caption{Different values of the electric field strength, $E$, used to get different excited state populations in the onsite model, $\varepsilon=\infty$ }
    \label{tab:my_label}
\end{table}

\begin{table}[h]
    \centering
    \begin{tabular}{||c||c|c|c|c|c|c|c|c||}
    \hline
         &$E_1$&$E_2$&$E_3$&$E_4$&$E_5$&$E_6$&$E_7$&$E_8$\\
         \hline
         $J_c=0.2$& 0.02& 0.08& 0.11&0.13&0.15&0.17&0.19&0.21\\
         \hline
         $J_c=0.4$& 0.03& 0.08& 0.11&0.14&0.16&0.18&0.20&0.22\\
         \hline
         $J_c=0.6$& 0.03& 0.08& 0.12&0.14&0.17&0.19&0.21&0.23\\
         \hline
         $J_c=1.0$& 0.03& 0.09& 0.12&0.15&0.18&0.20&0.22&0.25\\
         \hline
         $J_c=1.4$& 0.03& 0.09& 0.13&0.16&0.19&0.21&0.24&0.26\\
         \hline
         $J_c=1.8$& 0.03& 0.09& 0.13&0.16&0.19&0.22&0.25&0.28\\
         \hline
    \end{tabular}
    \caption{Different values of the electric field strength, $E$, used to get different excited state populations in the long-range interacting model with $\varepsilon=10$ }
    \label{tab:my_label}
\end{table}

\begin{table}[h ]
    \centering
    \begin{tabular}{||c||c|c|c|c|c|c|c|c||}
    \hline
         &$E_1$&$E_2$&$E_3$&$E_4$&$E_5$&$E_6$&$E_7$&$E_8$\\
         \hline
         $J_c=0.2$& 0.02& 0.08& 0.11&0.13&0.15&0.17&0.20&0.23\\
         \hline
         $J_c=0.4$& 0.03& 0.08& 0.11&0.14&0.16&0.18&0.21&0.24\\
         \hline
         $J_c=0.6$& 0.03& 0.08& 0.12&0.14&0.17&0.19&0.22&0.25\\
         \hline
         $J_c=1.0$& 0.03& 0.09& 0.13&0.15&0.18&0.21&0.24&0.27\\
         \hline
         $J_c=1.4$& 0.03& 0.09& 0.13&0.16&0.19&0.22&0.25&0.30\\
         \hline
         $J_c=1.8$& 0.03& 0.09& 0.13&0.16&0.19&0.23&0.27&0.33\\
         \hline
    \end{tabular}
    \caption{Different values of the electric field strength, $E$, used to get different excited state populations in the long-range interacting model with $\varepsilon=5$ }
    \label{tab:my_label}
\end{table}
\section{Burstein–Moss shift}

In this section we describe how the Burstein–Moss shift is calculated for Fig. 2 in the main-text.  When electrons are excited across the gap k-points can have partial occupations meaning optical transitions to $k$-points are rarely fully dissallowed by Pauli blocking.  For this reason we approximate the change in the optical gap by the following expression,
\begin{equation}
    E_{\mathrm{opt}} \sim \left|\frac{\sum_{k}\int_{\mu}^\infty \omega \mathcal{A}^<_k(\omega)}{\sum_{k}\int_{\mu}^\infty \mathcal{A}^<_k(\omega)}\right| + \left|\frac{\sum_{k}\int^{-\infty}_\mu \omega \mathcal{A}^>_k(\omega)}{\sum_{k}\int^{-\infty}_\mu \mathcal{A}^>_k(\omega)}\right|.
\end{equation}
This is a statistical approximation to the optical gap that gives the expected energy weighted by the electrons (holes) above (below) the Fermi energy.  From this we use the following relationship to compute the Burstein–Moss shift.

\begin{equation}
    E_{\mathrm{opt}} = E_\mathrm{g}^{(0)} + \Delta_{\mathrm{BGR}} + \Delta_{\mathrm{BM}}
\end{equation}

\section{Linear response optical gap for increasing inter-band interaction strength}

\begin{figure}
    \centering
    \includegraphics[width=\linewidth]{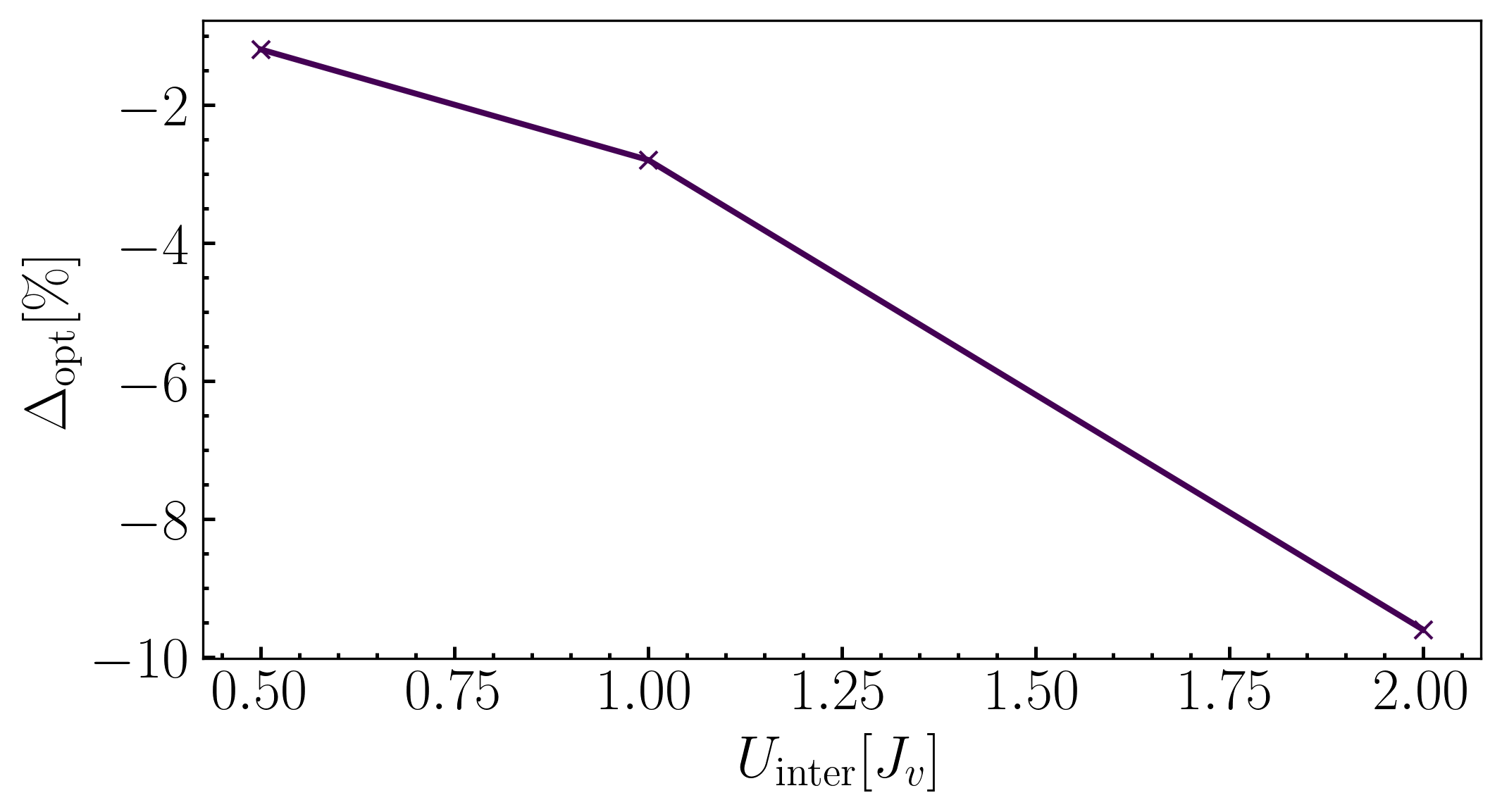}
    \caption{Difference between the gap in linear response regime ($N_c\rightarrow 0$) and the equilibrium fundamental gap for $U_\mathrm{inter}=0.5J_v$, $1.0J_v$ and $2.0J_v$.  Data shown for $J_c=J_v=1$.}
    \label{fig:opt_gap_U_inter}
\end{figure}

In Fig. \ref{fig:opt_gap_U_inter} we show how the linear response gap changes as we increase the interband coupling.  As we expect increasing the interband interaction strength leads to a larger renormalization of the gap even in the linear response regime.  This we interpret as coming from the increasing binding exciton binding energy which leads to a decrease of the optical gap. This result also helps explain the finite value of the optical gap as $N_c\rightarrow 0$ in Fig. 2 of the maintext.
\FloatBarrier
\FloatBarrier
\section{Sign of the renormalization in TD-HF}

The renormalization of TD-HF is dominated by the Hartree term where the difference in energies due to photo-excitation is (considering the Hartree contribution only),
\begin{equation*}
    \begin{split}
        h^\mathrm{H,c,\uparrow}_{ii} &\sim  U_\mathrm{intra}n^{c,\downarrow}_{i} +U_{\mathrm{inter}}n_{i}^{\mathrm{v}},\\
        &=n^{c,\downarrow}_{i}(U_{\mathrm{intra}}-2U_\mathrm{inter})
    \end{split}
\end{equation*}
for the conduction band and,
\begin{equation*}
    \begin{split}
        h^\mathrm{H,v,\uparrow}_{i} &\sim  U_\mathrm{intra}n^{v,\downarrow}_{i} +U_{\mathrm{inter}}n_{i}^{\mathrm{c}},\\
        &=n^{v,\downarrow}_{i}(U_{\mathrm{intra}}-2U_\mathrm{inter})
    \end{split}
\end{equation*}
for the valence band, where we only consider one spin due to symmetry.  Here we have used the relationship $n^v_i = 1-2n^{c,\downarrow}_i$ to simplify the expression. Since our system starts with $n_i^{v,\downarrow} = 1$ and $n_i^c=0$, upon photo excitation $n_i^v$ must decrease and $n_i^c$ must increase.  This means for $U_\mathrm{intra} > 2U_\mathrm{inter}$ the band gap will open after photo excitation and for $U_\mathrm{intra} < 2U_\mathrm{inter}$ the gap will close.  This explains the linear increase of the gap with increasing population in the conduction band as seen in Fig. 1 of the main text.  This argument does not include exchange effects however these do not strongly effect the renormalization and this is seen by the near perfect linear behavior of the renormalization.